\documentclass[journal,a4paper]{IEEEtran}  

\usepackage{cite}
\usepackage{amsmath,amssymb,amsfonts}
\usepackage[acronym,shortcuts]{glossaries}
\usepackage{graphicx}
\usepackage{color} 

\usepackage{lipsum}
\usepackage{graphicx}

\ifCLASSOPTIONcompsoc \usepackage[caption=false,font=normalsize,labelfont=sf,textfont=sf]{subfig}
\else
\usepackage[caption=false,font=footnotesize]{subfig}
\fi

\def\nTS{\ensuremath{T_\text{S}}}				 

\def\mathlette#1#2{{\mathchoice{\mbox{#1$\displaystyle #2$}}%
                           {\mbox{#1$\textstyle #2$}}%
                           {\mbox{#1$\scriptstyle #2$}}%
                           {\mbox{#1$\scriptscriptstyle #2$}}}} 
\renewcommand{\Vec}[1]{\mathlette{\boldmath}{#1}}  

\newcommand{\be}{\begin{equation}} 
\newcommand{\ee}{\end{equation}}
\newcommand{\ba}{\begin{array}}
\newcommand{\ea}{\end{array}}
\newcommand{\bdm}{\begin{displaymath}}
\newcommand{\edm}{\end{displaymath}}
\newcommand{\bea}{\begin{eqnarray}}
\newcommand{\eea}{\end{eqnarray}}
\newcommand{\bean}{\begin{eqnarray*}}
\newcommand{\eean}{\end{eqnarray*}}

\def\dif{\text{d}}

\def\nTS{\ensuremath{T_\text{S}}} 

\def\oT{\ensuremath{^\text{T}}} 

\begin{document}

\title{Site-Specific Radio Channel Representation for 5G and 6G}

\author{Thomas Zemen, Jorge Gomez-Ponce, Aniruddha Chandra, Michael Walter, Enes Aksoy, Ruisi He, David Matolak, Minseok Kim, Jun-ichi Takada, Sana Salous, Reinaldo Valenzuela, and Andreas F. Molisch
\thanks{T. Zemen (corresponding author, thomas.zemen@ait.ac.at) is with AIT Austrian Institute of Technology, his work is funded within the Principal Scientist grant Dependable Wireless 6G Communication Systems (DEDICATE 6G). J. Gomez-Ponce is with University of Southern California and 
ESPOL Polytechnic University, Guayaquil, Ecuador, he is supported by the National Science Foundation and the Foreign Fulbright Ecuador SENESCYT Program. A. Chandra is with NIT Durgapur, his work is funded by MeitY C2S program no. EE-9/2/2021-R\&D-E, GACR project no. 23-04304L, and NCN MubaMilWave no. 2021/43/I/ST7/03294. M. Walter is with the German Aerospace Center, E. Aksoy is with Fraunhofer Institute for Telecommunications, R. He is with Beijing Jiaotong University and his work is supported by the National Natural Science Foundation of China under Grant 62431003. D. Matolak is with University of South Carolina. M. Kim is with Niigata University and J. Takada is with Tokyo Institute of Technology; both are supported by the MIC, Japan, Grant JPJ000254. S. Salous is with Durham University, R. Valenzuela is with Nokia Bell Labs and A. F. Molisch is with University of Southern California, he is supported by the National Science Foundation.}
}

\maketitle

\newacronym{AI/ML}{AI/ML}{artificial intelligence/machine learning}
\newacronym{D-MIMO}{D-MIMO}{distributed multiple-input multiple-output}
\newacronym{DSD}{DSD}{Doppler spectral density}
\newacronym{FER}{FER}{frame error rate}
\newacronym{JCAS}{JCAS}{joint communication and sensing}
\newacronym{LIDAR}{LIDAR}{light detection and ranging}
\newacronym{LSF}{LSF}{local scattering function}
\newacronym{MIMO}{MIMO}{multiple-input multiple-output}
\newacronym{mmWave}{mmWave}{millimeter wave}
\newacronym{MPC}{MPC}{multipath component}
\newacronym{PDP}{PDP}{power delay profile} 
\newacronym{RIS}{RIS}{reconfigurable intelligent surface}
\newacronym{RMS}{RMS}{root mean square}
\newacronym{RX}{RX}{receiver}
\newacronym{SSCR}{SSCR}{site-specific radio channel representation}
\newacronym{TX}{TX}{transmitter}
\newacronym{HiL}{HiL}{hardware-in-the-loop}

\begin{abstract}
A site-specific radio channel representation (SSCR) takes the surroundings of the communication system into account by considering the environment geometry, including buildings, vegetation, and mobile objects with their material and surface properties. We present methods for an SSCR that is spatially consistent, such that mobile transmitter and receiver cause a correlated time-varying channel impulse response and closely spaced antennas are correctly correlated. An SSCR is composed of a dynamically varying number of multipath components solely defined by the environment geometry and the material of the environmental objects. Hence, the environment geometry is the only natural scenario parameterization and specific calibration procedures shall be avoided. 5G and 6G physical layer technologies are increasingly able to exploit the properties of a wide range of environments from dense urban areas to railways, road transportation, industrial automation, and unmanned aerial vehicles. The channel impulse response in this wide range of scenarios has generally non-stationary statistical properties, i.e., the Doppler spectrum, power delay profile, K-factor and spatial correlation are all spatially variant (or time-variant for mobile receivers). SSCRs will enable research and development of emerging 5G and 6G technologies such as distributed multiple-input multiple-output systems, reconfigurable intelligent surfaces, multi-band communication, and joint communication and sensing. We highlight the state of the art and summarize research directions for future work towards an SSCR.
\end{abstract}

\section{Introduction} 
A \ac{SSCR} allows the accurate prediction of radio propagation to analyze the performance of communication and sensing systems for sites of general interest and for different vertical domains such as railways, road transport, manned and unmanned aircraft or robotic collaboration. In this paper, we are interested in reviewing models that can influence research, system design, and standardization in different domains. 

Radio waves are emitted by the transmitting antenna and propagate via different propagation paths to the receiving antenna where they sum up, causing dispersion in delay and Doppler. The resulting radio channel can be fully described by a sampled double directional time-varying impulse response. The impulse response depends on the \ac{TX} and \ac{RX} coordinates, their velocity vectors as well as all objects in the environment. Its effect on the transmitted signal is computed by a time-varying convolution. 

Maxwell's equations can be solved numerically to obtain the impulse response; however, the numerical complexity is too high for most practical cases. Therefore, an \ac{SSCR} must identify a balance between precision and computational burden. The accuracy of an \ac{SSCR} improves with a more comprehensive geometry database and the incorporation of additional propagation effects, such as the reflection order, diffuse surface reflection, and diffraction effects. This paper will elucidate the currently known methods and algorithms to reduce the computational complexity of SSCR, achieving even real-time operation, while still aligning with empirical measurement data.

\subsection*{Organization of the paper}
In Section \ref{se:Motivation} we will motivate the need for an \ac{SSCR} that is able to cover research and development work for upcoming new physical layer technologies as described in Sec. \ref{se:PhyLayer} and new application scenarios as explained in Sec. \ref{se:ApplScen}. Three basic methods for an \ac{SSCR} are introduced in Sec. \ref{se:ModelType}, which use the scenario geometry as natural parameterization and avoids additional calibration steps. In Sec. \ref{se:StepByStep} we summarize a step-by-step description for the setup, validation, and use of an \ac{SSCR}. Open research directions for model validation techniques are introduced in Sec. \ref{se:ModelValidation} and unexplored channel measurement scenarios are outlined in Sec. \ref{se:EmpiricalMeasurements}. In Sec. \ref{se:Emulation} real-time radio channel emulation method for an \ac{SSCR} are explained. Final conclusions are drawn and open research directions are summarized in Sec. \ref{se:Conclusion}.

\section{Motivation for a site-specific channel model}
\label{se:Motivation}
The current channel model for the new radio (NR) physical layer of 5G and 6G is defined in \cite{3GPP38901}. It provides methods for link and system level simulations focusing on a small set of scenarios such as urban microcell, street canyon, urban macrocell, indoor office, rural macrocell, and indoor factory. The methods defined by 3GPP in \cite{3GPP38901} have a long history starting with GSM in the 1990s where a narrowband voice transmission channel could be well modeled using a power delay profile and a Doppler spectrum per channel tap. 

Over the last thirty years the channel modeling framework has been expanded to handle new physical layer features from 3G to 5G, such as wideband transmission, \ac{MIMO} techniques, and new frequency bands. The current version of \cite{3GPP38901} allows for geometry based modeling but considers only the first and last interaction point and the delay of a multipath component is not defined by the geometry. To achieve spatial and frequency consistency, specific algorithms and parameters are provided for each scenario. Clearly \cite{3GPP38901} aims at computational efficiency enabling link and system level result comparisons.

With upcoming new physical layer technologies such as \ac{JCAS}, \ac{D-MIMO} systems and \acp{RIS} a more comprehensive \ac{SSCR} is required, that (a) includes scenarios with non-stationary propagation properties such as vehicles on railways, streets, and in the air and (b) also allows for a wide aperture of antenna system (i.e., the near-field extends to multiple $\text{km}^2$). 

Hence, an \ac{SSCR} aims at a fundamental numerical modeling approach that is geometry based and can leverage the computing power available on modern graphical processing uni platforms as well as data driven methods. An \ac{SSCR} uses a dynamically varying number of \acp{MPC} solely defined by the environment geometry and the materials of the objects in the environment. An \ac{SSCR} shall inherently provide spatial and frequency consistent channel impulse responses, taking the geometry and the material properties of the environment objects into account without the definition of special procedures and parameters for each scenario. 

The long term work on an \ac{SSCR} is performed in the P1944 Standard working group under the aegis of the Mobile Communication Networks Standards Committee. It is important to distinguish the aims of P1944 from the channel modeling of 3GPP, 802.11/WiFi, and similar standards efforts. The latter aim to develop channel models best suited for the specific operating parameters and use cases of a standard. In contrast, P1944 aims to develop more generic models suitable for both academic research and industry developments, which might be used by standards organizations as basis for versions that are specialized to their purposes. 

\section{New physical layer technologies}
\label{se:PhyLayer}
Channel estimation is a fundamental challenge for any coherent communication system. The channel estimation algorithm needs to work reliably in all propagation scenarios. For the research and development of channel estimation methods, a link-level simulation with a realistic radio channel model is a key pre-requisite. For 5G and 6G technologies an \ac{SSCR} becomes a fundamental requirement to enable research and development in the areas described below. 

\subsection{Joint communication and sensing}
\begin{figure}
\centering
	\includegraphics[width=\columnwidth]{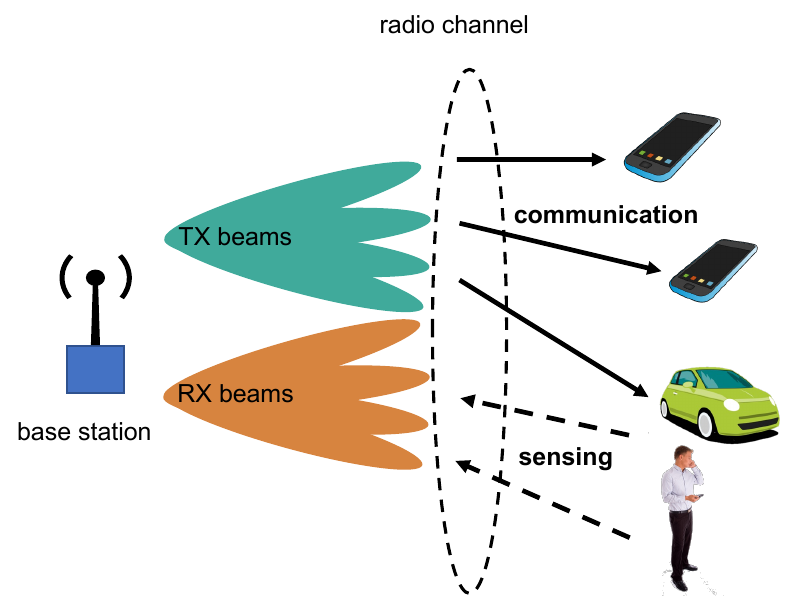} 
	\caption{Monostatic sensing.} 
\label{fig:Monostatic}
\end{figure}

Sensing capabilities will create a disruptive and explosive growth of new 5G and 6G applications and services. These can be economically implemented at minimal or no additional cost in existing base stations and/or user equipment heralding a new 5G and 6G era of \ac{JCAS} (also known as integrated sensing and communication). This sensing capability will span a very wide range of applications, including but not limited to: Intrusion detection, health monitoring, environmental monitoring, road traffic, daily life, industrial automation, etc. For example, a base station detects and alerts a pedestrian entering a pedestrian crossing while a car is about to enter, as shown in Fig. \ref{fig:Monostatic}.

Mono-static and bi-static network topologies are the simplest configuration for sensing. In mono-static sensing, a single base station implements the \ac{TX} and \ac{RX} functions at the same site, performing the sensing function within its radio coverage. In the bi-static case, two sites work together to achieve enhanced performance. The communication waveform may also be used opportunistically for the sensing function.
The reliable and robust design of such solutions will require new propagation models that enable deployment in a variety of environments for different applications with a high probability of user satisfaction. Current radio channel models are excellent in the context of a \ac{TX}-\ac{RX} link. However, at a minimum, the return power ratios from the object being sensed and its surrounding background are critical enhancements to be added to the existing propagation modes. Other improvements required are to include spatial consistency and Doppler statistics. 

\subsection{Antenna element group with wide aperture}
With emerging technologies such as \ac{RIS} and \ac{D-MIMO}, antenna elements are distributed over a large area. For \acp{RIS} potentially the whole facade of a building can be covered by $\lambda/2$ spaced antenna elements with adjustable reflection coefficients. For \ac{D-MIMO}, distributed remote radio heads can be deployed over an area with the size of about one $\text{km}^2$, being connected to a central processing unit (preferably with fibre optic links) that processes their signals jointly and coherently. In both cases an \ac{SSCR} is required that models the near field characterized by spherical wave fronts.

\subsection{Multi-band communication}
With decreasing wave-length of the \ac{TX} the received power decreases  quadratically. Hence, directive antennas are required for frequency range (FR) 2 ($24.25$ GHz to $52.6$ GHz) and FR3 ($7$ GHz to $24$ GHz) in 5G and 6G systems. The required beam search to establish a stable link between \ac{TX} and \ac{RX} is a challenging and time-consuming task in outdoor mobile scenarios. Hence, multi-band communication aims to use a radio signal, e.g. in FR1 ($0$ to $7$ GHz), to identify a suitable direction of arrival and direction of departure. These directions are used to establish a high-data rate broadband link in FR2 or FR3. For this technology a frequency-consistent \ac{SSCR} is needed.

\section{Application scenarios}
\label{se:ApplScen}
For the reliable communication with vehicles to enable remote control and safety applications, the wireless communication system must be able to handle non-stationary propagation conditions for the drive-by case. Such non-stationary transitions are not available in classic models such as \cite{3GPP38901} and require an \ac{SSCR}. Below we explore these future use cases in more detail and highlight scenario specific aspect that needs to be reflected by an \ac{SSCR}.

\subsection{Railways}
Railway communications must evolve from 2G (GSM rail, GSM-R) to the next generation railway communication system with improved performance and intelligence. 5G and related technologies \cite{he20225g} are seen as a solution to support the increasing data traffic, various new services, and high safety requirements for future intelligent railways. The next generation railway communication system is expected to support speeds of up to 500 km/h, which poses significant challenges for dynamic channel modeling, including dynamic estimation, tracking, and characterization of \acp{MPC}, which are not well studied for 2G railway communication systems and need to be addressed in the future. 
Channel dispersion in the delay, Doppler, and angle domain should also be accurately characterized by an \ac{SSCR} for future intelligent railways for a wide range of scenarios such as viaducts, tunnels, cuttings, stations, etc. For characteristic environments encountered in railway scenarios, an \ac{SSCR} demonstrates distinct responses tailored to each specific scene. For instance, in constrained environment like tunnels, it needs to account for the confined space and multiple reflections. In contrast, in open areas like viaducts, the focus might be on line-of-sight and diffraction over edges. Well-designed large and small scale models are essential for site-specific coverage prediction and link budget calculation. By tailoring specific methods and parameters to specific environments, an \ac{SSCR} can provide highly accurate and realistic simulations of radio wave propagation, enabling better design and optimization of wireless communication systems.

\subsection{Street transport}  
Vehicular communication on streets is defined by a wireless channel where potentially both, \ac{TX} and \ac{RX}, are moving. Hence, time-variant Doppler shifts play a major role for an \ac{SSCR} and scattering objects on both sides of the road, such as buildings, foliage, uneven terrain, etc. must be considered. Vehicular communication breaks down into different scenarios \cite{Bernado14}, ranging from static intra-vehicle or indoor parking to dynamic outdoor links between two vehicles or between a vehicle and the infrastructure. Vehicular radio channels are inherently location-specific resulting in a non-stationary fading process defined by the vehicle movement, and the movement of the surrounding vehicles in closer proximity. The building structure, street width and vegetation along the street have a strong influence on the radio channel properties as summarized in \cite{Bernado14}.

\subsection{Aircraft}
As the number of manned and unmanned aircraft continues to grow, there is a need to manage airspace more efficiently. This includes all aspects of communication, navigation and surveillance technologies. For all three aspects, it is important to understand the propagation channel between aircraft (manned or unmanned) and ground infrastructure. There are several key differences between the air-ground channel and traditional terrestrial communication systems. Aircraft can travel faster than terrestrial vehicles, both within and over areas where terrestrial communications take place. This poses challenges for accurate modeling of the air-ground channel, as the \ac{DSD} \cite{Walter20} and \ac{PDP} can be rapidly time-varying. 

Site-specific air-ground channel models are required, e.g., for small aircraft conducting \textit{fly-by} inspections of utilities such as large wind farms, advanced air mobility aircraft maneuvering near rooftops in urban areas, and air-ground communications with a large swarm of aircraft. In addition, \textit{aircraft-specific} effects such as airframe shadowing are not addressed in typical terrestrial channel modeling. 

In the common case where aircraft antennas are omnidirectional, the primary components of the channel impulse response are the line of sight and the earth surface reflection. The strength of the received line-of-sight and surface reflection also depends on the altitude of the aircraft, its bank angle or pitch during take-off and landing. These components must be supplemented by additional \acp{MPC}, which may be intermittent \acp{MPC} as the aircraft moves through a volume of space. In environments where the geometric and electrical properties of obstacles (or \textit{interacting objects}) can be accurately quantified, these intermittent \acp{MPC} can be estimated and tracked in the channel model. In the case of propeller aircraft, the radio signal is additionally Doppler-shifted by the frequency of the revolutions of the rotating propeller. 

\section{Methods for site specific channel representation}
\label{se:ModelType}
A radio channel can be fully described by a sampled double directional frequency response \cite{Steinbauer01}, which depends on the location of \ac{TX}, \ac{RX}, and the geometry of all relevant objects in the surrounding environment. Furthermore, material properties of all objects and their movement influence the double directional channel frequency response. It can be expressed as the sum of $P_m$ propagation paths.
\begin{multline}
g_{m,q}(\Vec{\alpha},\Vec{\beta})= \gamma_q \sum_{p=1}^{P_m}\eta_{p,m} e^{-j 2 \pi \theta_{p,m} q} \\
\delta(\Vec{\beta} - \Vec{\beta}_{p,m}) \delta(\Vec{\alpha} - \Vec{\alpha}_{p,m} ),
\label{eq:DD}
\end{multline}
where $m$ is the time index sampled at time $\nTS$, $q\in\{-Q/2,\ldots, Q/2-1\}$ is the frequency index and $Q$ is the even number of samples in the frequency domain, $\Vec{\alpha}_{p,m}=[\phi_m,\theta_m]\oT$ denotes the direction of arrival vector in terms of azimuth $\phi_m$ and elevation $\theta_m$ of \ac{MPC} $p$ at the \ac{TX} side and $\Vec{\beta}_{p,m}$ the corresponding direction of departure at the \ac{RX} side, respectively. The combined band-limited impulse response of the \ac{TX} and \ac{RX} hardware is denoted by $\gamma_q$. The normalized path delay is denoted by $\theta_{p,m}=\tau_{p,m}/(\nTS Q)$, the path weight by $\eta_{p,m}$, and $\tau_{p,m}$ is the delay of path $p$ at time index $m$.

The channel frequency response at the base point of the antennas, is finally obtained by integrating over the unit sphere surfaces $\mathcal{T}$ and $\mathcal{R}$ at \ac{TX} and \ac{RX} side, taking the antenna pattern of \ac{TX}, $\xi_\text{T}(\Vec{\alpha})$, and \ac{RX}, $\xi_\text{R}(\Vec{\beta})$, into account
\be
g_{m,q}=\oint_\mathcal{R} \oint_\mathcal{T} g_{m,q}(\Vec{\alpha},\Vec{\beta}) \xi_\text{R}(\Vec{\beta}) \xi_\text{T}(\Vec{\alpha}) \, \dif \Vec{\beta}\, \dif \Vec{\alpha}\,.
\label{eq:CTF}
\ee

The channel impulse response is related to the channel frequency response by the discrete inverse Fourier transform. The effect of the radio channel on the transmitted signal is computed by a time-varying convolution. 

\acp{SSCR} consider non-stationary radio propagation conditions. In all practical scenarios the radio channel exhibits a local wide-sense stationarity for a region in time and frequency \cite{Bernado14}. This property allows to simplify the \ac{SSCR} such that the \ac{MPC} parameters are constant for a stationarity region. The computation of the \ac{MPC} parameters can be done with different site-specific methods, as described below. These methods aim on general radio channel representation avoiding scenario specific parameterization or calibration procedures.

\subsection{Ray tracing}
Ray tracing is a high-frequency approximation used to solve Maxwell's equations. It has long been used successfully for deterministic channel prediction \cite{Valenzuela93}, particularly for deployment planning, both indoor and outdoor. As it computes the parameters of the \acp{MPC}, including amplitude, direction of arrival, and direction of departure, it provides a true double-directional channel characterization that can be combined with (almost) arbitrary antenna patterns and array patterns in post-processing. In addition, ray tracing results are inherently spatially consistent, i.e., the change in power, angles, and delays follows the environment's geometry, and - in a static environment - the same user equipment location is associated with the same channel. 

\begin{figure}
\centering
\includegraphics[width=0.9\columnwidth]{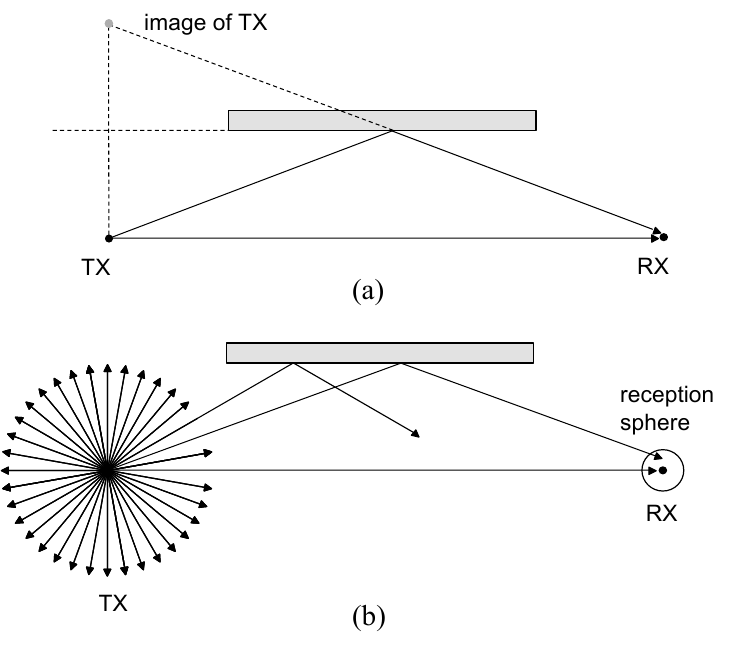}
\caption{(a) Image theory based ray tracer. (b) Ray launching.}
\label{fig:Raytracer}
\end{figure}

Ray tracing is typically implemented either as (i) image-theory-based ray tracing, where potential locations of image sources according to different orders of reflection are computed and rays between these locations and the \ac{RX} determined, or (ii) ray launching, where a \ac{TX} sends rays into different directions and follows their interactions with environmental objects until they arrive at the \ac{RX}, leave the area of interest, or become too weak to be of interest, see Fig. \ref{fig:Raytracer}. The former method is more efficient for point-to-point simulations, while the latter has advantages for many \ac{RX} locations and/or area coverage investigations. 

As the size of antenna arrays increases and wavelength decreases, wavefront curvature must be considered. Wavefront curvature implies that the \ac{RX} is within the Rayleigh distance of the \ac{TX}; as an important consequence, spatial multiplexing or mode multiplexing (for orbital angular momentum) becomes feasible. At the same time, operating under the assumption of homogeneous plane waves, when in reality, wavefront curvature occurs, can lead to a model mismatch in channel estimation, e.g., based on the sparsity assumption. Since the rays used in ray tracing are generally homogeneous plane waves, new ways of incorporating wavefront curvature are being explored. One way is to increase the number of rays representing a small angular range to satisfy the plane-wave assumption {\em locally}; however, this significantly increases the computational effort. 
\ac{D-MIMO} systems and ultra-massive \ac{MIMO} arrays are susceptible to another effect: different parts of the antenna elements/distributed radio units experience different amounts of shadowing. This effect, observed in several measurement campaigns, significantly impacts both channel capacity and transceiver implementations, such as sub-array selection in hybrid beamforming. Addressing this effect requires finer angular resolution, but this comes at the expense of efficiency.  

Evaluation at higher frequencies first requires more accurate and higher-resolution databases. These can be obtained, for example, from \ac{LIDAR} scanning of the environment or \ac{AI/ML}-based mapping of photographs onto full maps. In either case, the result is a point cloud, often with a resolution in the centimeter range. This leads to a dramatic increase in computation time for both image-based ray tracers and ray launchers. 

Open research areas for ways to increase efficiency include:
\begin{enumerate}
\item Visibility matrix-based methods, where a pre-processing step determines which surface elements (aka tiles) can "see" each other, so that no time is wasted tracking rays that cannot reach the \ac{RX} anyway.  
\item Appropriate choice of grid size, i.e., different parts of the point cloud can be merged into a single "effective" tile; such merging may depend on both the frequency considered and the distance of the tile from \ac{TX} and \ac{RX}. 
\item Treating complicated structures as a single "effective" structure: for example, instead of scanning and modeling each leaf of a tree, the tree is treated as a large structure that has an absorption coefficient and particular backscattering characteristics from its surface.
\item Dynamic ray tracing for mobile scenarios: Two state of the art methods for this purpose are a frame based approach, dividing the simulated time frame into different snapshots or the usage of velocity vectors for the moving objects within the scenario, to simulate the resulting ray paths at different points in time. Dynamic ray tracing can reduce complexity by (a) omitting a certain number of frames during the ray tracing simulations and interpolating the channel parameters in between; (b) predicting the resulting ray paths, based on the trajectory of the moving objects in the scenario, without additional ray tracing simulations. 
\end{enumerate}

Another research topic for high-frequency ray tracing is accurate models for diffuse scattering, as this phenomenon becomes more important with increasing frequency. Both, phenomenological approaches and models derived from first principles, have been investigated \cite{Jansen11_diffRT}. 

Finally, a ray tracer needs a calibration step. While the geometrical positions of environmental objects can be accurately determined, this is not the case for the electromagnetic properties of materials. The permittivity and conductivity of walls, windows, etc., can vary widely and are often determined by comparing the results of measurements at sample locations with the output of the ray tracer; the resulting coefficients are then used for all buildings in the area. While trial and error were typical for this task, more recently, differentiable ray tracers have been introduced, which allow more systematic optimization \cite{Hoydis23}.

\subsection{Quasi-deterministic method}
\label{sec:qdcm}
Quasi-deterministic \acp{SSCR}, also called geometry based stochastic channel models (GSCMs) in the literature, combine deterministic predictions, e.g., from a ray tracer, with stochastic but still geometry based channel modeling. 

Simplified ray tracing (a small number of reflections considered) is performed on a low-resolution environmental map to deterministically find the dominant MPC (the map can be actual or synthetically generated). Smaller \acp{MPC} associated with the dominant contributions are then generated stochastically either (i) in the geometric space by placing point scatterers according to a given distribution close to surfaces of geometric objects to represent surface roughness and point clusters to represent other objects such as vegetation, moving vehicles, and pedestrian or (ii) in the delay and direction of departure/direction of arrival domain. In both cases birth/death statistics of the associated \acp{MPC} and intra-cluster statistics are usually obtained from measurements. 

Quasi-deterministic models were first proposed by Kunisch and Pamp and by COST 259 in the early 2000s, and their concepts were then re-used by the European METIS and MiWeBa projects and adopted as an option for 3GPP simulations \cite{Weiler2016}. Compared to purely stochastic models (such as 3GPP), quasi-deterministic models have the advantages of spatial consistency (since the evolution of angles and delays of clusters or dominant paths follows from the geometry) and easier incorporation of specific geometries (e.g., the impact of street width can be easily incorporated). The advantages compared to ray tracing are the significantly reduced computational effort, easier exchange of environmental databases between entities that want to compare their results, and a straightforward way of incorporating mobile scatterers. 

\subsection{AI/ML based method}  
A promising approach of \ac{SSCR} is data-driven \ac{AI/ML}-based channel modeling \cite{Huang22}. \ac{AI/ML}-based channel models are trained to learn the mapping relationship between complex environmental features and channel characteristics. It can achieve a balance between generalization and prediction accuracy compared to traditional channel models. An \ac{SSCR} requires the prediction of channel characteristics of a specific site according to environmental conditions, and for \ac{AI/ML}-based modeling, the ability to extract features from environmental information determines the prediction accuracy of the model. 

For the extraction of environmental features, different data sources can be used, such as images, satellite maps, point cloud data, etc. The impact of data resolution on the accuracy of environmental features should be carefully considered. Methods like neural networks are usually used to extract propagation-related features from the data, which are further employed to predict channel characteristics. In the phase of model selection and network training, supervised learning is the most common approach, where models are trained to minimize prediction errors using labeled data, unsupervised learning methods are also employed in certain scenarios. For example, clustering algorithms can classify different channel conditions, and auto-encoders can extract meaningful features from raw data without labels. Additionally, generative adversarial networks can generate synthetic channel data, which is useful in scenarios with limited real-world data.

The developed AI network should take the environmental features as input and provide the corresponding site-specific channel parameters as output. Of course, physical radio propagation mechanisms and empirical channel models can be used and incorporated into the designed AI model to improve the site-specific channel prediction. Methods such as a regularization strategy, an early stop strategy and ensemble learning can also be used to avoid model over-fitting for a specific scenario. In addition, the ability of the model to self-evolve for different sites should be used as an evaluation index, and its essence lies in strengthening the existing model with additional data and features. Sufficient multi-source data and prior knowledge can support \ac{AI/ML}-based channel modeling for \ac{SSCR} needs with low complexity. 

The combined use of the three methods presented above is a promising future research direction to achieve the best trade-off between complexity and accuracy.

\section{Step by step description for \ac{SSCR} setup and validation}
\label{se:StepByStep}

Here we describe the steps needed to setup a \ac{SSCR} and use it for numerical link level simulation or for real-time channel emulation. The description illustrates the procedure for the quasi deterministic method for a vehicular scenario in detail but also indicates the basic procedure for the two other methods and other scenarios.

\begin{enumerate}
\item \emph{Environment definition:} Obtain a basic environment geometry from a database such as, e.g. open street map. Augment map data with objects detected via \ac{LIDAR} and video data. Figure \ref{fig:GSCM} shows an exemplary quasi-deterministic environment model for a vehicular scenario including buildings, as well as additional vegetation, parked cars, and city lamp posts. This additional objects improve the channel modeling accuracy substantially in terms of time-variant \ac{PDP} and time-variant \ac{DSD} as depicted in \cite[Fig. 5]{Dakic24}.
\begin{figure}
\centering
	\includegraphics[width=\columnwidth]{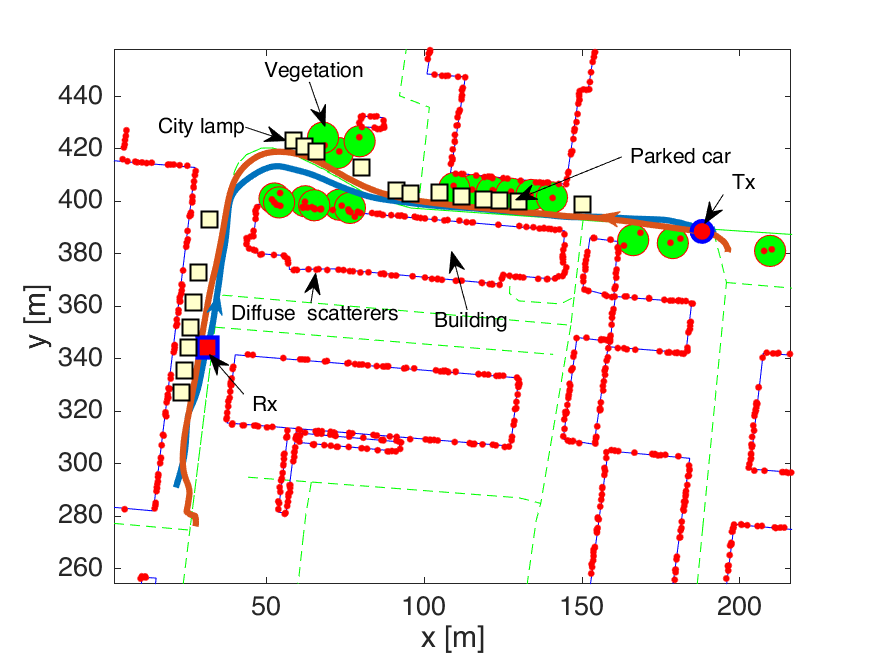} 
    \caption{Quasi-deterministic environment model for a vehicular scenario, see \cite[Fig. 4b]{Dakic24}. Red dots indicate point scatterers randomly distributed along walls and vegetation to represent static diffuse scattering, mobile discrete scatterers are represented by blue squares and circles, black squares represent static discrete scatterers such as parked cars or street lamps.} 
\label{fig:GSCM}
\end{figure}

\item \emph{Evaluate:} 
\begin{itemize}
\item Ray tracing: Use geometrical optics to compute line-of-sight, reflection up to a specified order $n$, diffraction and diffuses scattering \acp{MPC}.

\item Quasi deterministic: Distribute point scatterers in the geometry domain (shown as red dots in Fig. \ref{fig:GSCM}) along surfaces of objects according to a given distribution and compute \acp{MPC} between \ac{TX} and \ac{RX} taking blocking by buildings into account (alternatively scatterers can be also placed in the delay/angle domain). Use the \ac{MPC} path weight $\eta_{p,m}$ to limit the number of relevant \acp{MPC} $P_{m}$ for each stationarity region.

\item AI/ML: (i) Train the \ac{AI/ML} model with environment data and measured channel transfer functions for given \ac{TX} and \ac{RX} coordinates. (ii) Infer \acp{MPC} from the \ac{AI/ML} model based on new coordinates of \ac{TX} and \ac{RX}.
\end{itemize}

\item \emph{Use/validate model numerically:} Use the \ac{SSCR} by numerical evaluation of \eqref{eq:DD} and \eqref{eq:CTF} for link-level simulation or validate the \ac{SSCR} with empirical radio channel measurements using the methods described in Sec. \ref{se:ModelValidation}.

\item \emph{Real-time emulation:} Use the channel model for real-time emulation employing a \ac{HiL} setup, see Section \ref{se:Emulation} for more details. Real-time validation can be performed, e.g., comparing the frame error rate obtained in the \ac{HiL} setup with the one obtained by empirical measurements. In Fig. \ref{fig:FER_GSCM} (reproduced from \cite[Fig. 6]{Dakic24}) the \ac{FER} measured on the road is compared with the \ac{FER} obtained by emulating the channel transfer function from the quasi deterministic \ac{SSCR}. Two cases are shown: (i) the case when all identified objects are included, and (ii) the case when only objects from open street map are included. A close match in terms of \ac{FER} between measurement and \ac{SSCR} based emulation can be observed in Fig. \ref{fig:FER_GSCM} for the former case (i). 
\begin{figure}[ht]
	\begin{center}
		\includegraphics[width=1.0\columnwidth]{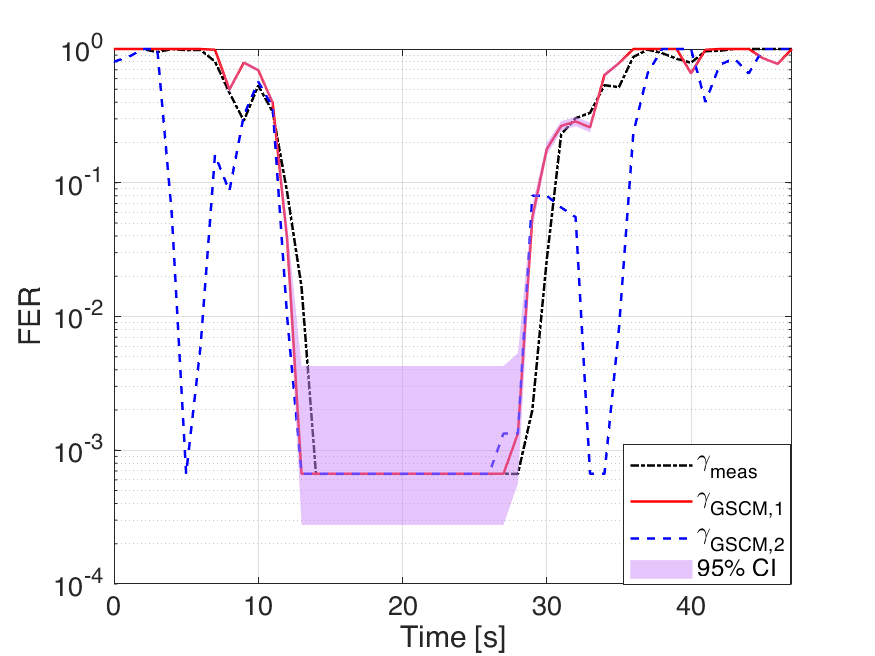}
	\end{center}
	\caption{\ac{FER} comparison reproduced from \cite[Fig. 6]{Dakic24}: \ac{FER} $\gamma_{\text{meas}}$ obtained directly on the road, \ac{FER} $\gamma_{\text{GSCM,1}}$ obtained by emulating channel transfer function from the GSCM in the case when all identified objects are included, and \ac{FER} $\gamma_{\text{GSCM,2}}$ obtained by emulating CTF from the GSCM in the case when only objects from open street map are included. The violet region indicates the 95\% confidence interval (CI) of the binomially distributed \ac{FER}.}
	\label{fig:FER_GSCM} 
\end{figure}
\end{enumerate}

\section{Model validation}
\label{se:ModelValidation}
An \Ac{SSCR} shall be able to represent non-stationary fading conditions while being frequency and spatially (i.e. time) consistent for the evaluation of 5G and 6G technologies that use large antenna apertures, perform multi-band operation and implement \ac{JCAS} algorithms. To validate an \ac{SSCR} we need to establish respective validation methods by opening new research directions, such as:
\subsubsection{Non-stationary environment evaluation}
\label{subsec:NonStationary}
The \ac{LSF} \cite{Bernado14} $\hat{\mathcal{C}}_{s;n,r}$ can be used to evaluate non-stationary measurement data and compare it with the \ac{SSCR} outcome. With the time-variant frequency response estimate $\hat{g}_{m,q}$, the \ac{LSF} can be computed as
\begin{multline}
\hat{\mathcal{C}}_{s;n,r}=\frac{1}{IJ}\sum_{w=0}^{IJ-1}\Bigg\vert \sum_{m'=-\frac{M}{2}}^{\frac{M}{2}-1} \sum_{q=-\frac{Q}{2}}^{\frac{Q}{2}-1} \hat{g}_{m'+Ms,q} \cdot  \\
G_{w;m,q}\mathrm{e}^{-\mathrm{j}2\pi(rm-nq)} 
\Bigg\vert^2,
\label{eq:LSF}
\end{multline}
with the Doppler index $r\in \{-M/2\ldots, M/2-1\}$, the delay index $n\in\{0,\ldots, Q-1\}$, and the tapers $G_{w;m, q}$ are two-dimensional discrete prolate spheroidal sequences as shown in detail in \cite{Bernado14}. The number of tapers $IJ$ controls the bias-variance trade-off of the \ac{LSF} estimate. Using the \ac{LSF} $\hat{\mathcal{C}}_{s; n,r}$ we can compute the time-variant \ac{PDP}, \ac{DSD} and path loss as marginals with respect to Doppler, time or both.

\subsubsection{Frequency domain continuity}
Another validation requirement for \ac{SSCR} is frequency domain continuity, i.e. ensuring the applicability of channel parameters over the entire frequency range of operation. 5G NR pushes the traditional sub-$6$ GHz boundary to $7.125$ GHz (FR$1$) and introduces \ac{mmWave} bands in FR$2$. FR$3$ is called the upper mid-band and is being considered for 5G-Advanced and 6G. Experiments in the sub-THz range from $52.6$-$114.25$ GHz (FR$4$) and $114.25$-$275$ GHz (FR$5$), as well as in the THz range ($>300$ GHz) have attracted interest. The validation of parameters over such different frequency ranges prompted a series of simultaneous multi-band measurements. The blocking and deviation characteristics of a scatterer become more pronounced as the carrier frequency increases. This means that small scatterers, which could be safely ignored for FR$1$ or FR$3$ bands, may need to be included in the \ac{SSCR} when applied to bands for FR$2$ or above. 

\subsubsection{Backscatter radio channel properties}
\Ac{JCAS} requires new radio channel properties to be represented by an \ac{SSCR}. Hence, radar cross-section parameters need to be included for path loss calculation in mono-/bi-static sensing. If the link lengths for \ac{TX}-target and target-\ac{RX} are given by $d_1$ and $d_2$, then the path loss is given by $\gamma_{\mathrm{dB}}(d_1,d_2,\sigma,f) = \gamma_{\mathrm{dB}}(d_1) + \gamma_{\mathrm{dB}}(d_2) + 10\log\frac{\lambda^2}{4\pi}- 10\log(\sigma)$, where $\sigma$ is the normalized reflectivity of the target. Measuring the radar cross section of different targets and classifying them according to their radar cross section range are important objectives for \ac{JCAS} field test campaigns. Additional \ac{SSCR} validation parameters include clutter/scattering patterns and object mobility, depending on the accuracy, resolution, and refresh rate requirements of the \ac{JCAS} use case. The return power from natural or man made backgrounds is also critical for defining the target to noise power ratio.

\section{Empirical radio channel measurements}
\label{se:EmpiricalMeasurements}
An integral part of site-specific channel model development is model validation through extensive field measurements. Channel parameters characterized by traditional channel sounding measurements can be loosely grouped into three categories: large-scale (over an area representative of urban, sub-urban, or residential environment), medium scale (over a path or small area such as a square, within the same room or residential dwelling) and small scale (over a short distance typically a few wavelengths). 

\begin{figure}
\centering
	\includegraphics[width=\columnwidth]{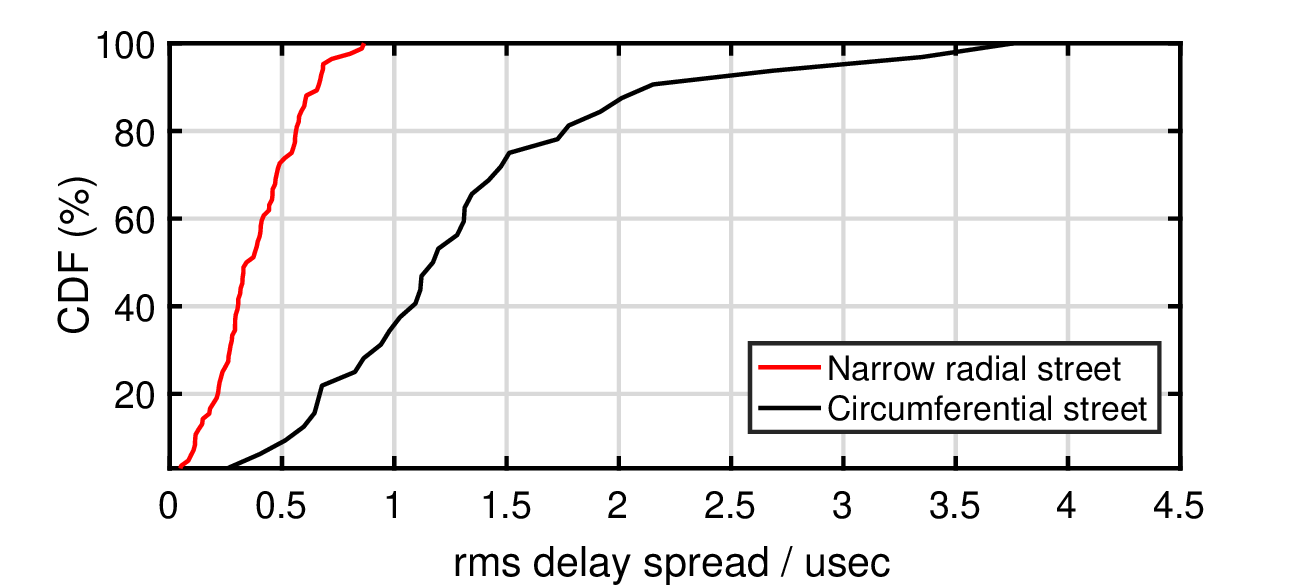}
	\caption{\ac{RMS} delay spread over two medium scale routes in the city of Manchester at 2.1\,GHz. The red curve show the \ac{RMS} delay spread cumulative distribution function for a road radially oriented towards the base station and the black line depicts data from a circumferential street with respect to the base station location  \cite{Salous07}.} 
\label{fig:RMSurban}
\end{figure}

In earlier models pathloss, delay spread and Doppler spread were modeled as function of the environment.
\acp{SSCR} aim to include also their dependence on distance as well as on the surrounding environment and its geometrical properties. As an example we depict the \ac{RMS} delay spread cumulative distribution function for two medium scale urban routes that have different orientations with respect to the TX (in Manchester, UK, at 2.1\,GHz) in Fig. \ref{fig:RMSurban}. 

New 5G and 6G inspired use cases, as described in Section II, require new empirical radio channel measurements for verification, which constitute interesting future research directions:
\subsubsection{Vehicular scenarios}
For rail and road applications, new measurement campaigns are aimed at capturing the interplay between higher centre frequency and highly mobile site-specific features. Challenges for such campaigns include ensuring repeatable and fault-free operation of the channel sounder under extreme environmental conditions (weather, vibration, etc.) and maintaining proper synchronization between different subsystems. The node mobility causes fluctuations in the beamwidth, and as a result, the number of \acp{MPC} varies proportionally with the beamwidth. The channel becomes highly non-stationary, and a local scattering function, as described in subsection \ref{subsec:NonStationary} and \cite{Bernado14}, can be used to characterize and compare the measured and modelled channel. 

\subsubsection{Airborne scenarios}
For airborne scenarios, the mobile-to-mobile (M2M) channel model validation requires at least two additional facets: first, consideration of the full three-dimensional site geometry (unlike vehicular scenarios where a 2D model is generally sufficient), and second, consideration of scattering components (along with line-of-sight and specular reflections). The validation is done in the delay-Doppler plane to account for the fast mobility patterns of \ac{TX} and \ac{RX}. Classical site-general parameters, such as shadowing variance, are augmented with new features like shadowing correlation distances to ensure the validity of an airborne mobile-to-mobile channel model at different \ac{TX}/\ac{RX} heights.

\subsubsection{Urban cellular high frequency scenarios}
For example, \cite{Tsukada_2024} builds on the quasi deterministic channel model type described in Sec. \ref{sec:qdcm}. The authors present a recipe for \ac{mmWave} MPC generation for a specific urban cellular environment in Yokohama, Japan. Here, the measured data is used to estimate an exponential decay model to calibrate the ray tracer. Random clusters are generated with the measured site-specific statistical parameters of large and small scale fading parameters. This step improves the accuracy of the quasi-deterministic channel model by incorporating real-world measurement data that captures the specific characteristics of the environment under consideration.

\section{Channel emulation}
\label{se:Emulation}
A site-specific representation (digital twin) for a radio communication channel relies on a radio channel emulator to carry out hardware-in-the-loop tests. Transmit and receive modems are connected to the radio channel emulator to repeatedly test the communication system under well-defined propagation conditions. Channel emulators for dynamic non-stationary site-specific channel models including live objects are currently not available on the commercial market. 

The challenges to realize such an emulator is the broadband communication link between the numerical channel model and the convolution unit for the sampled impulse response. The required data rate $R \sim c_{1} B^2 T_{\text{D}}$ increases quadratically with the bandwidth of the communication system $B=1/\nTS$, where $c_{1}$ is a constant describing the number of bits per complex sample and $T_{\text{D}}$ is the support of the delay spread.

\begin{figure}
\centering
  \includegraphics[width=\columnwidth]{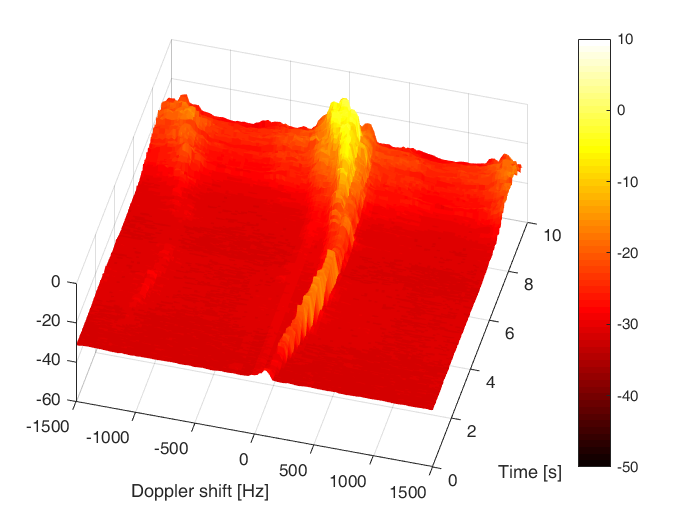}
	\caption{Time variant \ac{DSD} obtained from an emulator controlled by an \ac{SSCR} (measured with a channel sounder), see \cite[Fig. 13b]{Hofer19}.} 
\label{fig:tvDSD}
\end{figure}

In \cite{Hofer19} a novel site-specific channel emulation approach is presented where the convolution is approximated with a reduced rank subspace representation of dimension $D\ll M$ using discrete prolate spheroidal sequences. This approach that is fundamentally \ac{MPC} based, avoids the quadratic increase of $R$ with $B$ and enables site-specific channel emulation. In Fig. \ref{fig:tvDSD} we reproduce \cite[Fig. 13c]{Hofer19}, showing the time-variant \ac{PDP} from the subspace based radio channel emulator that is controlled by an \ac{SSCR}. A good match with the measured radio propagation scenario in \cite[Fig. 13b]{Hofer19} is achieved.
Combining the subspace based channel emulation methods with a quasi deterministic \ac{SSCR} allows for real-time hardware-in-the-loop emulation \cite{Dakic24}.

Another approach in \cite{Keerativoranan24} divides the environment into a predefined grid and pre-calculates the propagation paths for each grid point via ray tracing. For a mobile device moving within a grid, complex exponential interpolation is performed between adjacent grid points to achieve a continuous channel response.

\section{Conclusions and future research}
\label{se:Conclusion}   
In this paper we have reviewed the motivations for including \acp{SSCR} in future standards (e.g. 5G, 6G and WiFi 7). Ray tracing, quasi-deterministic and \ac{AI/ML}-based methods were introduced and their specific advantages and disadvantages for \acp{SSCR} were explored. Model validation methods for non-stationary environments, frequency domain continuity, and backscatter radio channel characteristics were explained. Empirical radio channel measurements for validation of path loss characteristics, urban cellular, vehicular, and airborne scenarios are reviewed. Finally, low-complexity channel emulation techniques for hardware-in-the-loop testing of communications hardware using an \ac{SSCR} are presented.

Future research directions for \acp{SSCR} are (a) acquisition of empirical measurement data for vehicular and airborne scenarios, (b) methods to compute \acp{MPC} that achieve a good tradeoff between accuracy and complexity, (c) new model validation techniques, and (d) real-time channel emulation methods that directly exploit the \ac{MPC}-based architecture of \acp{SSCR}.

\Acp{SSCR} are an emerging field with great importance for 5G and 6G wireless communication systems to support research and development of emerging \ac{D-MIMO}, \ac{RIS}, multi-band communication, and \ac{JCAS} techniques. 

\section*{Acknowledgement}
We would like to thank COST INTERACT for helpful discussion.



\begin{thebibliography}{10}
\providecommand{\url}[1]{#1}
\csname url@samestyle\endcsname
\providecommand{\newblock}{\relax}
\providecommand{\bibinfo}[2]{#2}
\providecommand{\BIBentrySTDinterwordspacing}{\spaceskip=0pt\relax}
\providecommand{\BIBentryALTinterwordstretchfactor}{4}
\providecommand{\BIBentryALTinterwordspacing}{\spaceskip=\fontdimen2\font plus
\BIBentryALTinterwordstretchfactor\fontdimen3\font minus
  \fontdimen4\font\relax}
\providecommand{\BIBforeignlanguage}[2]{{%
\expandafter\ifx\csname l@#1\endcsname\relax
\typeout{** WARNING: IEEEtran.bst: No hyphenation pattern has been}%
\typeout{** loaded for the language `#1'. Using the pattern for}%
\typeout{** the default language instead.}%
\else
\language=\csname l@#1\endcsname
\fi
#2}}
\providecommand{\BIBdecl}{\relax}
\BIBdecl

\bibitem{3GPP38901}
3GPP, ``Study on channel model for frequencies from 0.5 to 100 {GHz. (Release
  17)},'' 3rd Generation Partnership Project (3GPP), Technical Report (TR)
  38.901, December 2023.

\bibitem{he20225g}
R.~He, B.~Ai, Z.~Zhong, M.~Yang, R.~Chen, J.~Ding, Z.~Ma, G.~Sun, and C.~Liu,
  ``5{G} for railways: Next generation railway dedicated communications,''
  \emph{{IEEE} Commun. Mag.}, vol.~60, no.~12, pp. 130--136, 2022.

\bibitem{Bernado14}
L.~Bernad{\'o}, T.~Zemen, F.~Tufvesson, A.~Molisch, and C.~Mecklenbr{\"a}uker,
  ``Delay and {Doppler} spreads of non-stationary vehicular channels for safety
  relevant scenarios,'' \emph{{IEEE} Trans. Veh. Technol.}, vol.~63, no.~1, pp.
  82 -- 93, 2014.

\bibitem{Walter20}
M.~Walter, D.~Shutin, M.~Schmidhammer, D.~W. Matolak, and A.~Zajic, ``Geometric
  analysis of the {Doppler} frequency for general non-stationary {3D}
  mobile-to-mobile channels based on prolate spheroidal coordinates,''
  \emph{{IEEE} Trans. Veh. Technol.}, vol.~69, no.~10, pp. 10\,419--10\,434,
  2020.

\bibitem{Steinbauer01}
M.~Steinbauer, A.~F. Molisch, and E.~Bonek, ``The double directional radio
  channel,'' \emph{{IEEE} Antennas Propag. Mag.}, vol.~43, no.~4, pp. 51--63,
  August 2001.

\bibitem{Valenzuela93}
R.~Valenzuela, ``A ray tracing approach to predicting indoor wireless
  transmission,'' in \emph{IEEE Veh. Techn. Conf. (VTC)}, Secaucus, NJ, USA,
  May 1993, pp. 214--218.

\bibitem{Jansen11_diffRT}
C.~Jansen, S.~Priebe, C.~Moller, M.~Jacob, H.~Dierke, M.~Koch, and T.~Kurner,
  ``Diffuse scattering from rough surfaces in {THz} communication channels,''
  \emph{IEEE Trans. Terahertz Science and Techn.}, vol.~1, no.~2, pp. 462--472,
  2011.

\bibitem{Hoydis23}
J.~Hoydis, F.~A. Aoudia, S.~Cammerer, M.~Nimier-David, N.~Binder, G.~Marcus,
  and A.~Keller, ``{Sionna RT: D}ifferentiable ray tracing for radio
  propagation modeling,'' \emph{arXiv preprint arXiv:2303.11103}, 2023.

\bibitem{Weiler2016}
R.~J. Weiler, M.~Peter, W.~Keusgen, A.~Maltsev, I.~Karls, A.~Pudeyev,
  I.~Bolotin, I.~Siaud, and A.-M. Ulmer-Moll, ``Quasi-deterministic
  millimeter-wave channel models in {MiWEBA},'' \emph{EURASIP Journal on
  Wireless Communications and Networking}, vol. 2016, no.~1, p.~84, Mar 2016.

\bibitem{Huang22}
C.~Huang, R.~He, B.~Ai, A.~F. Molisch, B.~K. Lau, K.~Haneda, B.~Liu, C.-X.
  Wang, M.~Yang, C.~Oestges \emph{et~al.}, ``Artificial intelligence enabled
  radio propagation for communications --- {Part II}: Scenario identification
  and channel modeling,'' \emph{{IEEE} Trans. Antennas Propag.}, vol.~70,
  no.~6, pp. 3955--3969, 2022.

\bibitem{Dakic24}
A.~Daki{\'c}, B.~Rainer, P.~Priller, G.~Nan, A.~Momi{\'c}, X.~Ye, and T.~Zemen,
  ``Wireless {V2X} communication testbed for connected, cooperative and
  automated mobility,'' in \emph{IEEE Veh. Networking Conf. (VNC)}, Kobe,
  Japan, May 2024.

\bibitem{Salous07}
S.~Salous and H.~Gokalp, ``Medium-and large-scale characterization of
  {UMTS}-allocated frequency division duplex channels,'' \emph{IEEE
  Transactions on Vehicular Technology}, vol.~56, no.~5, pp. 2831--2843, 2007.

\bibitem{Tsukada_2024}
\BIBentryALTinterwordspacing
H.~Tsukada, N.~Suzuki, B.~Bag, R.~Takahashi, and M.~Kim, ``Millimeter-wave
  urban celluar channel characterization and recipe for high-precision
  site-specific channel simulation,'' \emph{techRxiv}, Feb. 2024. [Online].
  Available: \url{http://dx.doi.org/10.36227/techrxiv.170831411.10437905/v1}
\BIBentrySTDinterwordspacing

\bibitem{Hofer19}
M.~Hofer, Z.~Xu, D.~Vlastaras, B.~Schrenk, D.~Loeschenbrand, F.~Tufvesson, and
  T.~Zemen, ``Real-time geometry-based wireless channel emulation,''
  \emph{{IEEE} Trans. Veh. Technol.}, vol.~68, no.~2, pp. 1631 -- 1645, July
  2019.

\bibitem{Keerativoranan24}
N.~Keerativoranan, K.~Saito, and J.~Takada, ``Grid-based channel modeling
  technique for scenario-specific wireless channel emulator based on path
  parameters interpolation,'' \emph{IEEE Open Journal of the Com. Soc.},
  vol.~5, pp. 1724--1739, 2024.

\end{thebibliography}


\IEEEbiographynophoto{Thomas Zemen}
is Principal Scientist at the AIT Austrian Institute of Technology and docent at TU Wien. He leads the reliable wireless communication group at AIT.
\vspace{-3mm}

\IEEEbiographynophoto{Jorge Gomez-Ponce}
is with the University of Southern California, Los Angeles, CA, USA. He is also with the ESPOL Polytechnic University, Guayaquil, Ecuador.
\vspace{-3mm}

\IEEEbiographynophoto{Aniruddha Chandra}
is an Associate Professor at NIT Durgapur and Secretary of the IEEE P2982 standard working group.
\vspace{-3mm}

\IEEEbiographynophoto{Michael Walter}
is member of the scientific staff at the Institute of Communications and Navigation at the German Aerospace Center (DLR).
\vspace{-3mm}

\IEEEbiographynophoto{Enes Aksoy}
is working towards his doctoral degree at Fraunhofer Institute for Telecommunications and the Technical University of Berlin, Germany..
\vspace{-3mm}

\IEEEbiographynophoto{Ruisi He}
is a Professor at Beijing Jiaotong University and leads the modern communication research institute.
\vspace{-3mm}

\IEEEbiographynophoto{David Matolak}
is Professor of Electrical Engineering at the University of South Carolina, and leads the Wireless Science and Engineering (WiSE) Laboratory.
\vspace{-3mm}

\IEEEbiographynophoto{Minseok Kim}
is an Associate Professor at Niigata University, Japan. His current focus is on 6G mmWave/THz communications.
\vspace{-3mm}

\IEEEbiographynophoto{Jun-ichi Takada}
is Dean and Professor, School of Environment and Society, Tokyo Institute of Technology.
\vspace{-3mm}

\IEEEbiographynophoto{Sana Salous}
holds the Chair in Communication Engineering at Durham University and is the Director of the Centre for Communication Systems.
\vspace{-3mm}

\IEEEbiographynophoto{Reinaldo Valenzuela,} Member NAE, IEEE and Nokia Bell Labs Fellow.
\vspace{-3mm}

\IEEEbiographynophoto{Andreas F. Molisch}
is the Golomb-Viterbi Chair Professor at the University of Southern California; he is Fellow of NAI, AAAS, IEEE, IET, and Member of Au. Ac. Sci.


\end{document}